\documentclass[a4paper,pra,preprint,floatfix,superscriptaddress]{revtex4}
\usepackage[english]{babel}
\usepackage{amsmath,amsfonts,amssymb,float}
\usepackage{graphicx}
\usepackage{dcolumn,bm}
\usepackage{verbatim} 
\usepackage{color}
\usepackage{url}
\usepackage{float}

\def\clf{Central Laser Facility, STFC Rutherford Appleton Laboratory, Didcot, OX
11 0QX, United Kingdom}
\def\strathclyde{Department of Physics, SUPA, University of Strathclyde, Glasgow, G4 0NG, United Kingdom}

\def\cockcroft{The Cockcroft Institute, Sci-Tech Daresbury, Warrington WA4 4AD, United Kingdom}

\def\N{\mathbb{N}}
\def\Z{\mathbb{Z}}

\renewcommand{\vec}[1]{\ensuremath{\mathbf{#1}}}

\begin{document}

\title{A tuneable frequency comb via dual-beam laser-solid harmonic generation}

\author{Raoul Trines}
\email[Corresponding author, ]{raoul.trines@stfc.ac.uk}
\author{Holger Schmitz}
\affiliation\clf
\author{Martin King}
\author{Paul McKenna}
\affiliation\strathclyde
\affiliation\cockcroft
\author{Robert Bingham}
\affiliation\clf
\affiliation\strathclyde
\date\today

\begin{abstract}
A high-power laser pulse at normal incidence onto a plane solid target will generate odd harmonics of its frequency. However, the spacing of the harmonic lines in this configuration is fixed. Here, we study harmonic generation using two laser beams incident on a plane target at small, opposite angles to the target normal, via particle-in-cell simulations. When looking at the harmonic radiation in a specific direction via a narrow slit or pinhole, we select an angle-dependent subset of the harmonic spectrum. This way, we obtain a harmonic frequency comb that we control via the observation angle and the input laser frequency. The spectra for s- and p-polarised harmonics are studied separately, as they offer different frequency combs. The divergence of the harmonic radiation will be reduced by using wider laser spots, thus increasing the efficacy of the scheme. We will discuss extensions to this scheme, such as using beams with unequal frequencies, a slight tilt of the target, or employing more than two beams.
\end{abstract}

\maketitle

\section{Introduction}


In ``classic'' high harmonic generation (HHG) by high-intensity laser beams in isotropic configurations, the frequency spectrum contains odd integer multiples of the frequency of the pump beam, e.g. for HHG in ionised gas \cite{ferray} or HHG via normal incidence onto a solid target \cite{lichters}. The wavelength of hte pump laser ranges from the mid- and near-infrared (IR) to the ultraviolet, which in turn changes the HHG spectrum from a coherent supercontinuum to a collection of narrowband peaks \cite{popmin12,popmin15}. However, greater control of the frequency spectrum, especially larger spacing between the peaks, is important for numerous applications, e.g. to tune the effective carrier frequency of the attosecond pulses created via HHG. Hence the concept of the ``frequency comb'', where only a subset of peaks from the harmonic spectrum is selected, not all peaks, see e.g. \cite{alon98,bayku21,rego22,trines24}. The specific subset of selected peaks is tuneable via higher order properties (e.g. transverse shape, polarisation or phase) of either the driving beams or the target used in the HHG process.

Recently, Rego \emph{et al.} \cite{rego22} developed the concept of the ``necklace beam''. In this configuration, two beams with parallel linear polarisations (LP) and equal frequencies but different orbital angular momentum (OAM) levels, no common divisor and not both odd, e.g. $\ell_A = +2$ and $\ell_B = -3$, are used to drive harmonics in an isotropic medium (gas). When the harmonic light is observed through an on-axis pinhole, only the light with zero OAM is caught; this light is found to contain only the harmonics $(2n+1)|\ell_A - \ell_B|$, so odd multiples of the frequency $|\ell_A - \ell_B| \omega_0$ rather than $\omega_0$. This is of course tuneable via $\ell_{A,B}$.

Asymmetric frequency combs have been considered using a single pump laser with circular polarisation (CP) interacting with a target with a finite rotational symmetry, see e.g. Refs. \cite{alon98,bayku21,trines24}. Again, the on-axis harmonic light is collected through a pinhole and only those harmonic multiples that have an OAM-free component  are caught.

In this paper, we show that the role of the OAM level $\ell$ in the ``necklace beam'' configuration can also be assumed by a small transverse wave vector component $k_\perp$. To this end, we consider two identical LP beams with frequency $\omega_0$ and s-polarisation, crossing at a small angle. This configuration is similar to that displayed in Figure 3a of Hickstein \emph{et al.} \cite{hickstein}. However, we turn the question around: rather than asking ``for a given frequency, which emission angles do we see'', we ask ``for a given emission angle, which frequencies do we see''. We predict that for a beam crossing half-angle $\alpha_0$ and a viewing angle $\alpha_v$ with $\tan\alpha_0 = N\tan\alpha_v\ (N\in\N)$, the harmonic spectrum contains peaks at $(2n+1)N\omega_0$. Thus, $N$ takes the role of $|\ell_A - \ell_B|$ in the necklace beam configuration. We note that $N$ is tuned via the observation angle $\alpha_1$, which is easier and more versatile than tuning via $|\ell_A - \ell_B|$. It also allows for more extensions, which will be discussed below.

HHG with non-collinear laser pulses has been studied in the past \cite{hickstein, fomichev, bertrand, heyl14, gariepy, zli, negro, ellis}. In most cases, the laser beams had different wave lengths, e.g. a fundamental at 800 nm and its second harmonic at 400 nm. While a 2-D $(\omega,k_\perp)$ spectrum is usually shown, in most  cases the angular diffraction pattern is studied for a given harmonic order. In our work, we aim to start from a given diffraction angle, and study the subset of harmonic orders that is diffracted in that specific angle. We also study the harmonic spectra for s- and p-polarised harmonics separately, as these offer different harmonic progressions and thus different options for frequency combs.

The divergence of the harmonic modes deserves special attention. For a Gaussian driving beam, the effective radius of the part of the envelope that generates a given harmonic, i.e. where the pulse intensity is above a given threshold, decreases rapidly with increasing harmonic order. As a result, higher order harmonics are generated by an ever smaller part of the pump beam(s), so their divergence increases with harmonic order. For a ``necklace beam'': the region where the harmonics are generated has a ring shape, so its radius $R_0$ will be roughly the same regardless of the harmonic order. The Rayleigh length of harmonic $q\omega_0$ is then given by $z_q = k_q R_0^2/2 = qk_0 R_0^2/2 = qz_0$, explaining why Rego \emph{et al.} \cite{rego22} find a decreasing divergence for. increasing harmonic order. In our work, we use driving pulses with super-Gaussian envelopes to ensure that the effective radius of the harmonic-generating region does not decrease too much for increasing harmonic order.

\section{Theory}

As in our previous work \cite{trines24}, we decompose all our light into pure CP modes with a single frequency $\omega > 0$, wave vector $\vec{k}$, OAM level $\ell$ and spin $\sigma = \pm 1$. We will then consider harmonic progressions in $(\omega/\sigma,\vec{k}/\sigma,\ell/\sigma)$ space. In this space, harmonic progressions are always regular grids, usually with one or two degrees of freedom.

For the original necklace beam \cite{rego22}, the laser configuration is given by the four points $\pm(1,\ell_{A,B})$ in $(\omega/\sigma,\ell/\sigma)$ space. The harmonic progression then has two degrees of freedom and is given by:
\[
(\omega/\sigma,\ell/\sigma)_{mn} = (1,\ell_A) + m(2, \ell_A + \ell_B) - n(0, \ell_A - \ell_B).
\]
On axis, modes with zero OAM will have an intensity maximum, while modes with nonzero OAM will have an intensity minimum. Selecting the on-axis light via a pinhole will thus preferentially select modes with zero OAM. Such modes are found for $\ell_A + m(\ell_A + \ell_B) - n(\ell_A - \ell_B) = 0$. For e.g. $\ell_A = +2$ and $\ell_B = -3$ (so $|\ell_A + \ell_B| = 1$), this yields $2-m-5n=0$, or $\omega/\sigma = 1 + 2(2-5n) = 5(1-2n)$, $n \in \Z$. On recombination, this yields the odd multiples of $5\omega_0$ found in Ref. \cite{rego22}. Note in particular that for a given value of $n$, the ``rows'' in the harmonic progression are not quite horizontal, e.g. $(\omega/\sigma,\ell/\sigma)_{m} = (1,2) + m(2,-1)$ for $n=0$. This ensures that each row will land one point on the horizontal $\ell/\sigma=0$ axis, leading to a peak in the spectrum of the on-axis harmonic light.

For a scenario with two identical s-polarised LP beams crossing at a shallow angle $2\alpha_0$ (we will use $\alpha_0 = 15^\circ$ in our simulations below), as used by e.g. Hickstein \emph{et al.} \cite{hickstein}, the vector potential $\vec{A}_0 \propto \vec{e}_s$ of the driving laser beams is given by the four points $(\pm 1,\pm k_{0\perp})$ in $(\omega/\sigma,k_\perp/\sigma)$ space, with $k_{0\perp} = k_0 \sin\alpha_0$. The lowest order term driving the s-polarised harmonics is then $A_0^2 \vec{A}_0$. The harmonic progression for the s-polarised light has two degrees of freedom and is given by:
\begin{equation}
\label{eq:progression}
(\omega/\sigma, k_\perp/\sigma)_{mn} = (1,1) + m(2,0) + n(0,2),
\end{equation}
 i.e. all rows are horizontal and there will be no harmonic light in the fully forward direction. However, the scenario of the necklace beam can be replicated by looking at the harmonic light at a slight angle with respect to the fully forward direction. We note that the angle between the ``rows'' and the axis for the necklace beam is determined by $\ell_{A,B}$; to change that angle one needs to change the experiment. However, the viewing angle in a scenario with two crossing beams is free, and multiple angles can be studied in the same experiment. This demonstrates the flexibility of our approach.

We also note, as already discussed by Trines \emph{et al.} \cite{trines24}, that every 2-D harmonic progression contains the possibility of a frequency comb. It is simply a matter of finding the correct viewing angle.

For our case, we note that the crossing half-angle of the pump beams is given by $\tan\alpha_0 = k_{\perp 0}/k_{\parallel 0}$, while the emission half angle of any harmonic mode is given by $\tan\alpha_{mn} = (2n+1)k_{\perp 0}/[(2m+1)k_{\parallel 0}]$. For e.g. a viewing angle $\alpha_v$ given by $\tan\alpha_v = k_{\perp 0}/(Nk_{\parallel 0})$, with $N$ an odd integer, we preferentially select modes $q\omega_0$ with $q = 2m+1 = (2n+1)N$, $n\in\N$. We compare this to the ``necklace beam'' scenario, where a frequency comb $q = 2m+1 = (2n+1)|\ell_A + \ell_B|$ is obtained \cite{rego22}. However, in our case the value of $N$ is tuned via the viewing angle $\alpha_v$, which is easier than tuning via the pump beam OAM levels $\ell_{A,B}$.

One can consider a generalisation to $ck_\perp/\omega = \tan(\alpha_0)N_\perp/N_\parallel$ with $\gcd(N_\perp, N_\parallel) = 1$. However, this does not yield different spectra. Since $q = 2m+1$ must be an odd integer, one does not get a different frequency comb for $N_\perp \not= 1$, e.g. $ck_\perp/(\omega/\tan(\alpha_0)) = 3/7$ instead of $1/7$. But in experiments, it may help to have multiple available angles to obtain a specific comb.

In addition to the s-polarised harmonics, we note that the work by Lichters \emph{et al.} \cite{lichters} also predicts p-polarised harmonics at even multiples of the base frequency and wave vector, even for purely s-polarised pump beams. These harmonics are emitted by the ``DC mode'' $\vec{A}_\mathrm{DC} = \tan(\alpha_0) \vec{e}_p$ via the nonlinear term $A_0^2 \vec{A}_\mathrm{DC}$. The harmonic progression for the p-polarised light also has two degrees of freedom and is given by
\begin{equation}
\label{eq:p-progression}
(\omega/\sigma, k_\perp/\sigma)_{mn} = m(2,0) + n(0,2).
\end{equation}
For these harmonics, the emission half angle of any harmonic mode is given by $\tan\alpha_{mn} = (2n)k_{\perp 0}/[(2m)k_{\parallel 0}] = (n/m) \tan\alpha_0$. For e.g. a viewing angle $\alpha_v$ given by $\tan\alpha_v = k_{\perp 0}/(Nk_{\parallel 0})$, with $N$ any integer, we preferentially select modes $q\omega_0$ with $q = 2m = (2n)N$, $n\in\N$. We note two differences with the s-polarised frequency combs give above: (i) any value for $N$ is now possible, not just odd ones, and (ii) the frequency of the lowest harmonic is now the same as the frequency step size, rather than half the step size.

\section{Simulation setup}

We carried out 2-D particle-in-cell simulations using the Epoch code \cite{epoch1,epoch2} to model the interaction of two laser beams hitting a flat target at near-normal incidence. Numerical parameters were: a box size of $16\times64\ \mu\mathrm{m}^2$, a grid size of $4096\times8192$ cells, 10 particles per cell, and a total simulation time of 250 fs. Physical parameters were: a laser wave length of 1 $\mu$m, a laser peak intensity of $3\cdot 10^{19}$ W/cm$^2$, a beam waist of 16 $\mu$m, a transverse pulse envelope that is either Gaussian ($\exp(-r^2)$) or super-Gaussian ($\exp(-r^{16})$), a Gaussian longitudinal envelope with a duration of 50 fs, a plasma slab with a maximum density of $8n_\mathrm{cr}$ and a pre-plasma layer extending 1 $\mu$m in front of the slab. The laser beams are identical with linear s-polarisation and impact the same spot at angles of $\pm \alpha_0 = \pm 15^\circ$ with respect to the target normal.

We run 2-D simulations because this is sufficient for this problem, it requires fewer resources and it allows us to use higher resolution, improving the quality of our Fourier spectra.

We need a fairly wide spot diameter: $k_0 R \tan(15^\circ) > 1$. Otherwise: (i) there will not be enough ``grating'' periods within the spot, so the harmonic amplitude becomes poor, (ii) the harmonics will have a large divergence angle, so harmonics from different emission angles overlap and selecting a specific viewing angle becomes difficult or impossible, (iii) the divergence angle for higher harmonics will be compromised  \cite{rego22}: a given harmonic needs a minimum intensity to be generated ``non-perturbatively'', so the spot area where this intensity is reached should be large enough to ensure a narrow divergence angle in the far field. In Ref. \cite{rego22} the pump laser spot is always a ring, so does not really get smaller for a higher intensity threshold, but in our case the spot diameter will get smaller for a higher intensity threshold, so one start with a decent spot diameter and laser power in the first place. The effects of this can also be seen in the poor harmonic divergence obtained for Gaussian driving beams, and the much improved divergence obtained for super-Gaussian beams (see below).

\begin{figure*}[hb]
        \centering
(a)  \includegraphics[width=0.6\textwidth]{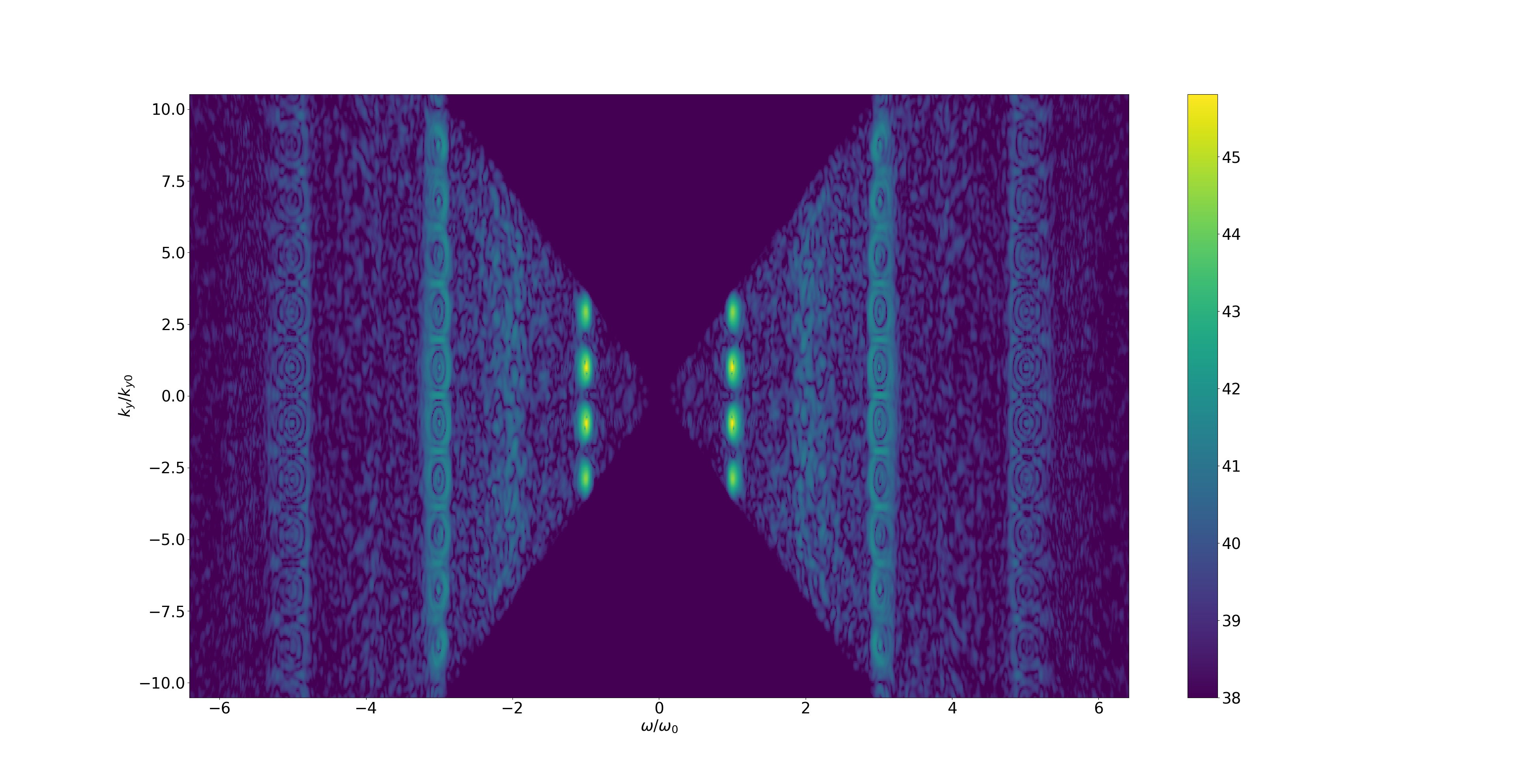}
(b)  \includegraphics[width=0.3\textwidth]{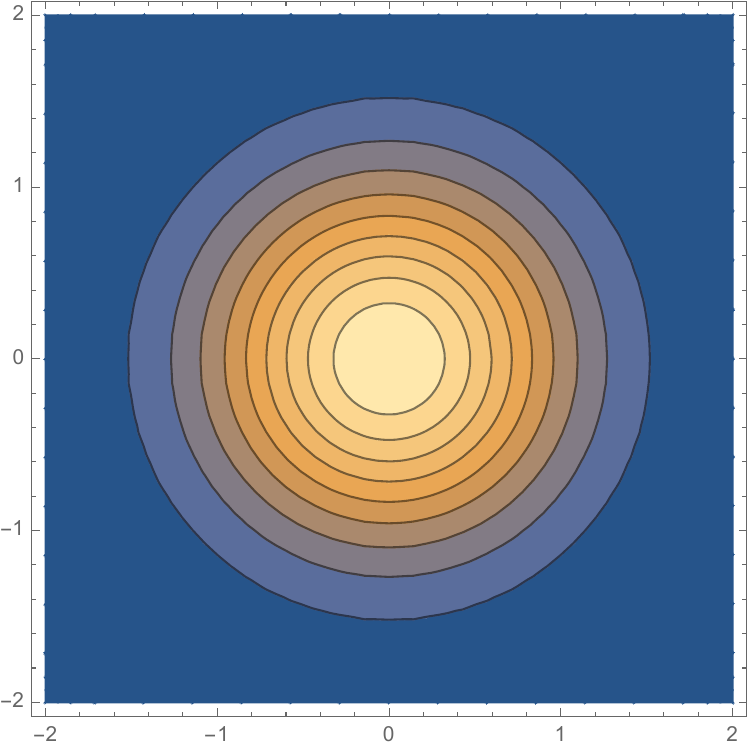}

(c)  \includegraphics[width=0.6\textwidth]{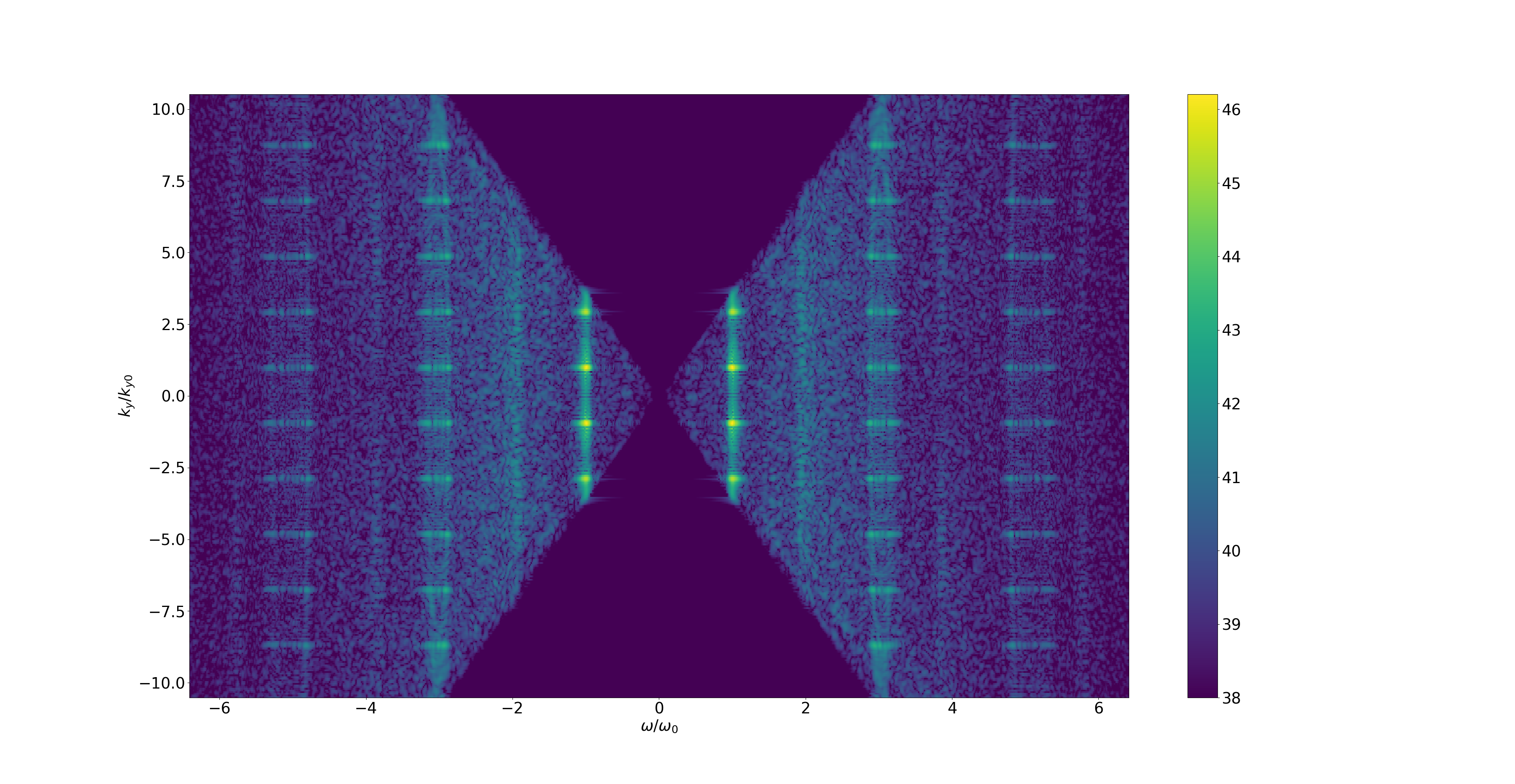}
(d)  \includegraphics[width=0.3\textwidth]{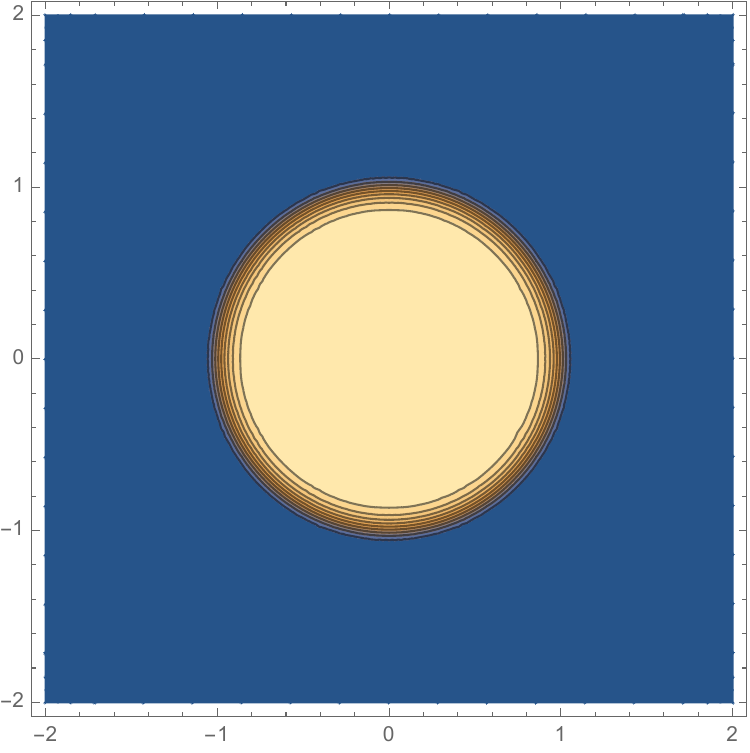}
        \caption{Comparison of  the quality of the s-polarised spectral peaks generated by either (a,b) two Gaussian beams or (c,d) two super-Gaussian beams. For the Gaussian beams, the divergence of the harmonic peaks is too large to be practical, see frame (a), while for the super-Gaussian beams, the divergence has much improved, see (c). This can be explained from the contours of the beam envelope: for a Gaussian intensity envelope, the radius of the contours decreases quickly for increasing energy, see (b), while for the super-Gaussian envelope, the contour radius remains much larger for increasing energy. Thus, in the case of a super-Gaussian beam, the harmonics are generated from a much larger ``effective spot diameter'' than for a Gaussian beam, leading to improved harmonic divergence. }
        \label{figure1}
\end{figure*}

\section{Simulation results}

\subsection{S-polarised harmonics}

\begin{figure*}[hb]
        \centering
        \includegraphics[width=\textwidth]{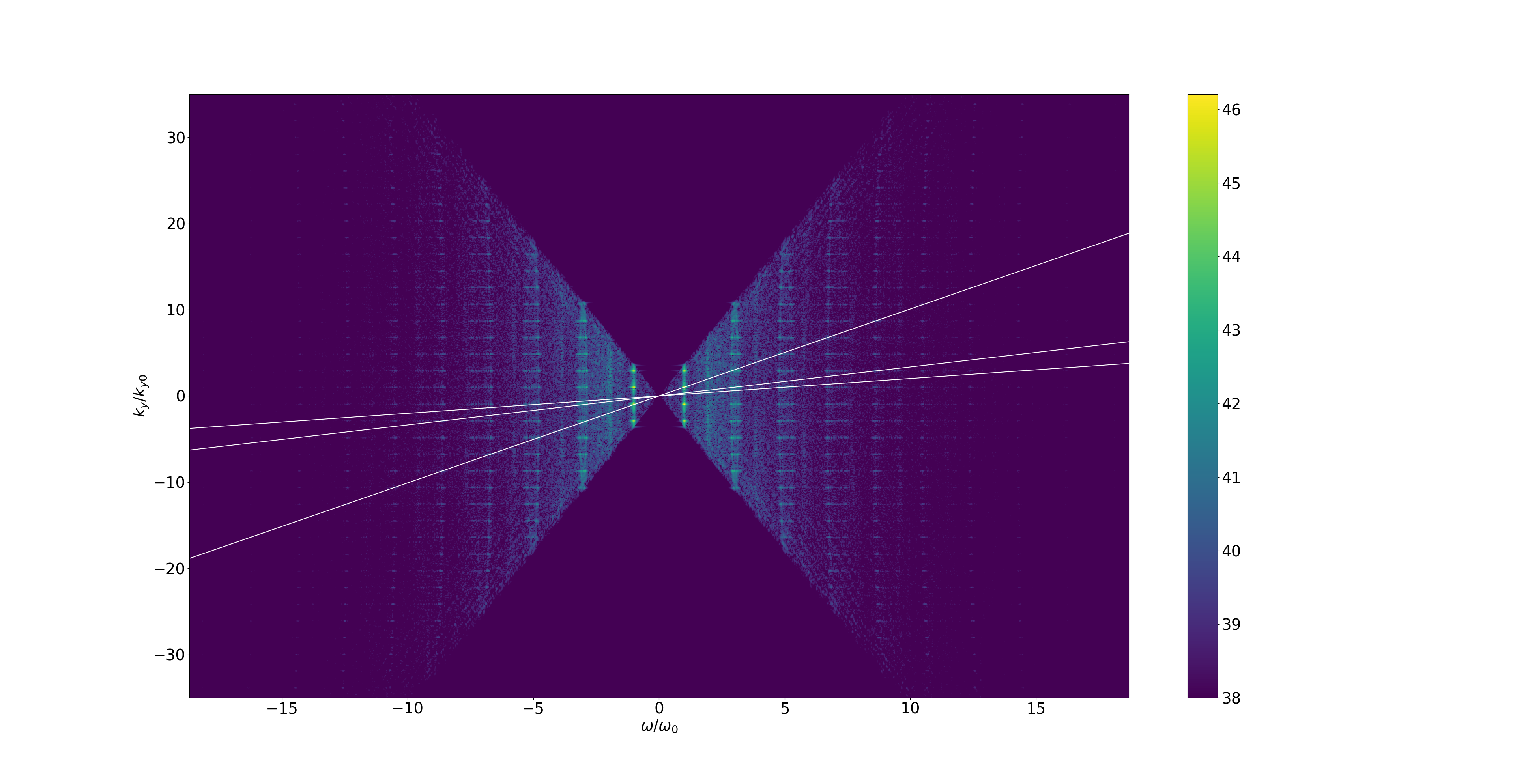}
        \caption{The $(\omega/\sigma, k_\perp/\sigma)$ spectrum of the reflected s-polarised harmonic radiation for the super-Gaussian beams. The positions of the peaks are as predicted by Eq. (\ref{eq:progression}). Line-outs are taken at $ck_\perp/\omega = \tan(15^\circ)/N$ for $N=1, 3, 5$, as indicated in the plot (smaller slope means higher $N$). }
        \label{figure2}
\end{figure*}

\begin{figure*}[ht]
(a)    \includegraphics[width=0.4\textwidth]{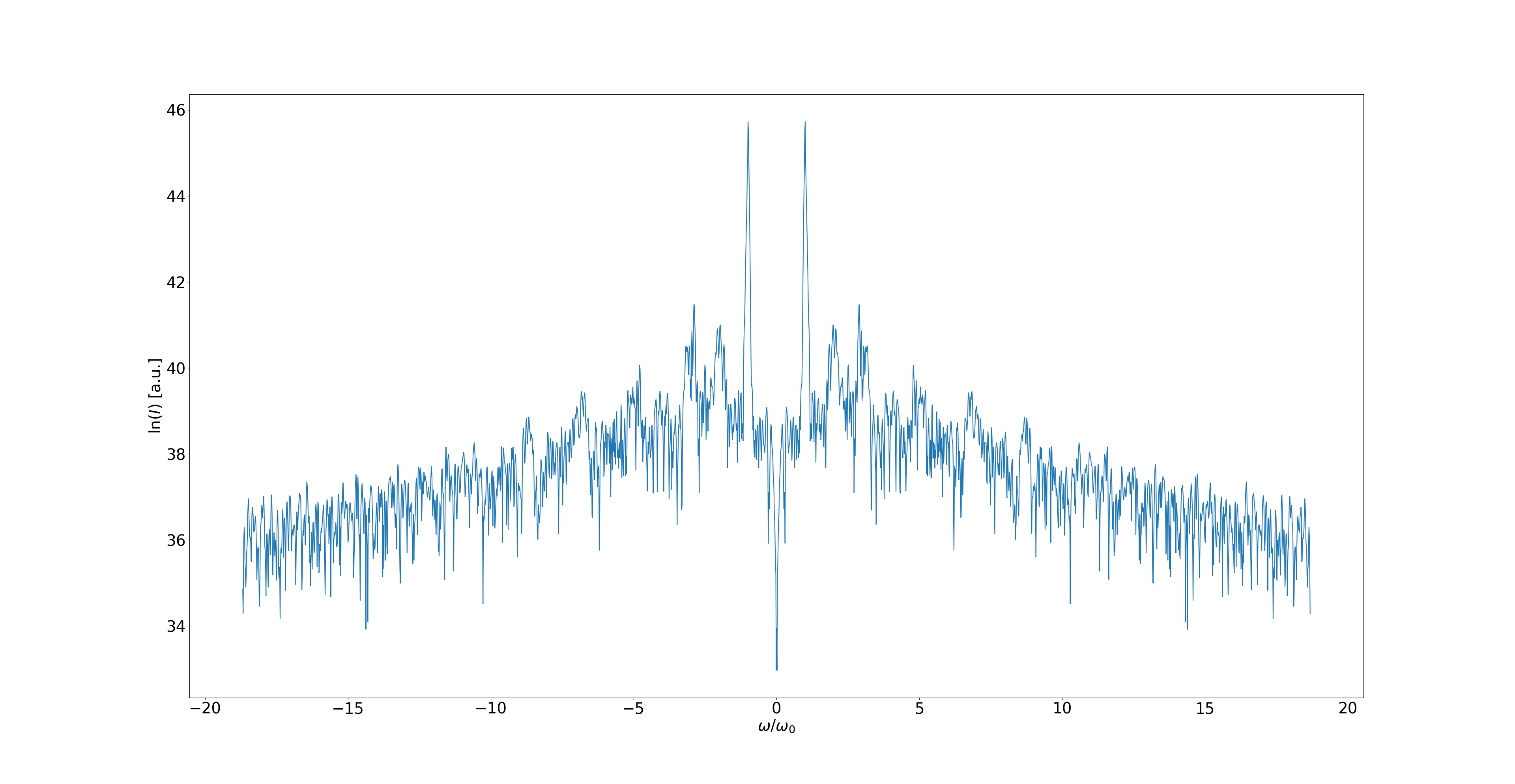}
(b)    \includegraphics[width=0.4\textwidth]{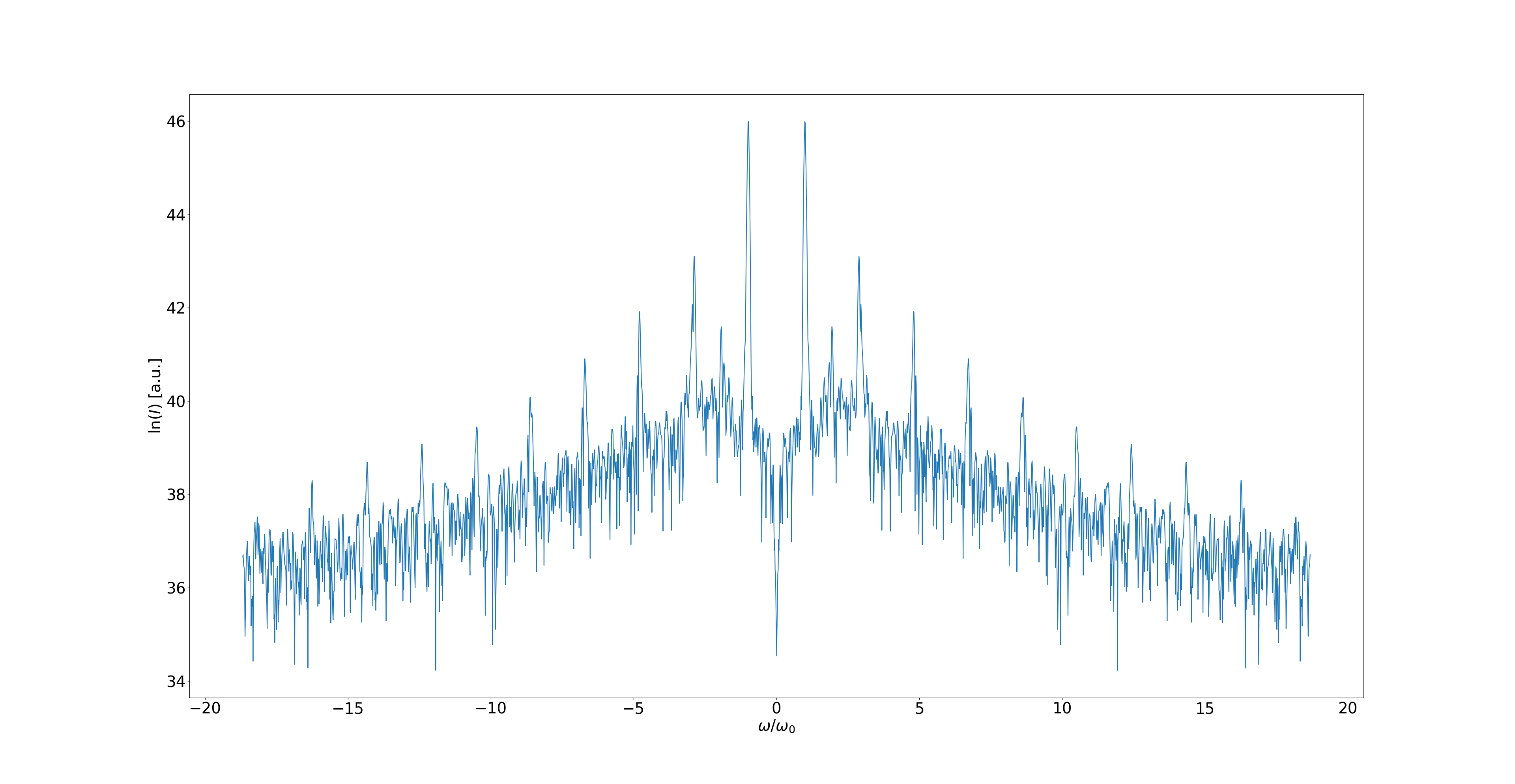}

(c)    \includegraphics[width=0.4\textwidth]{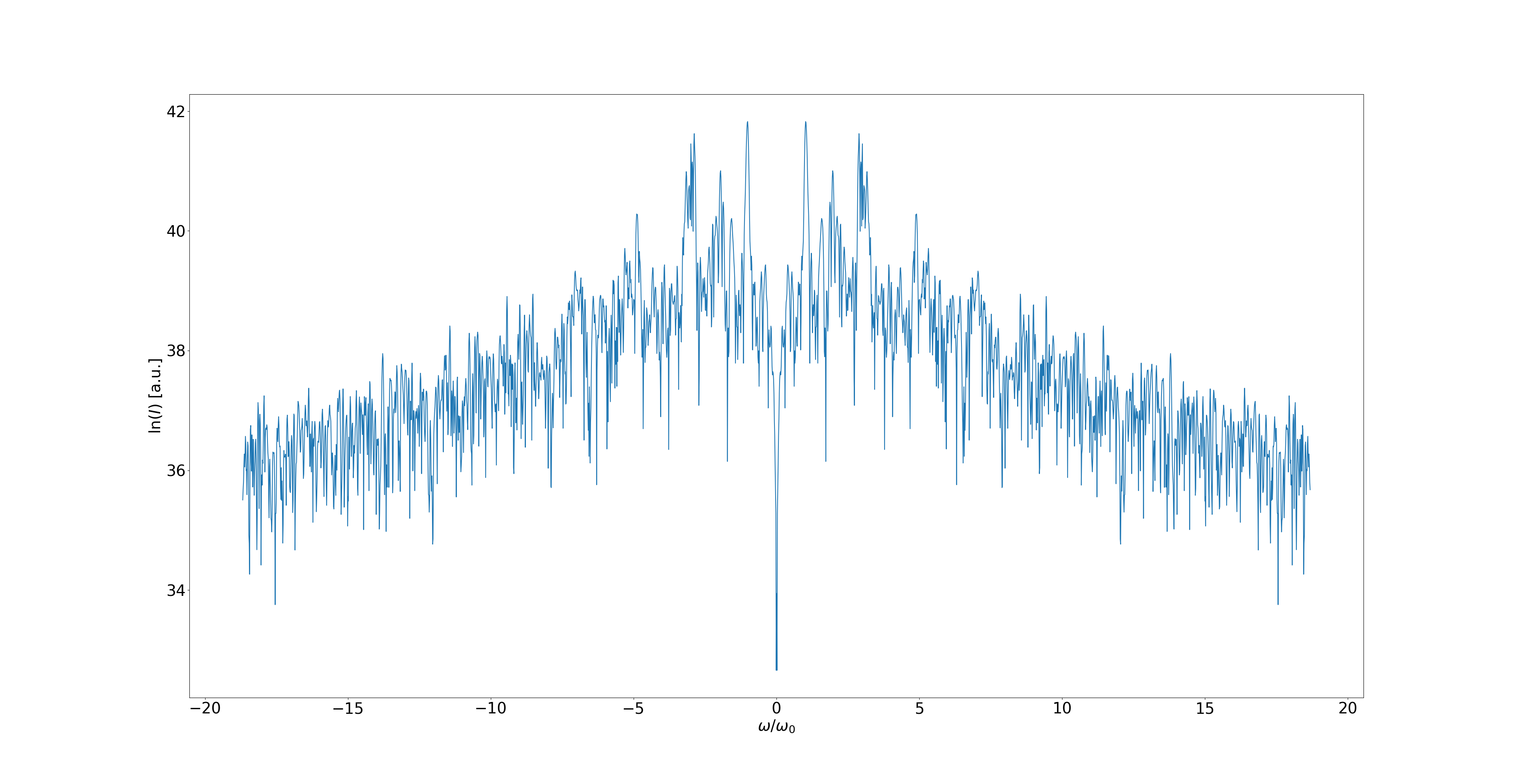}
(d)    \includegraphics[width=0.4\textwidth]{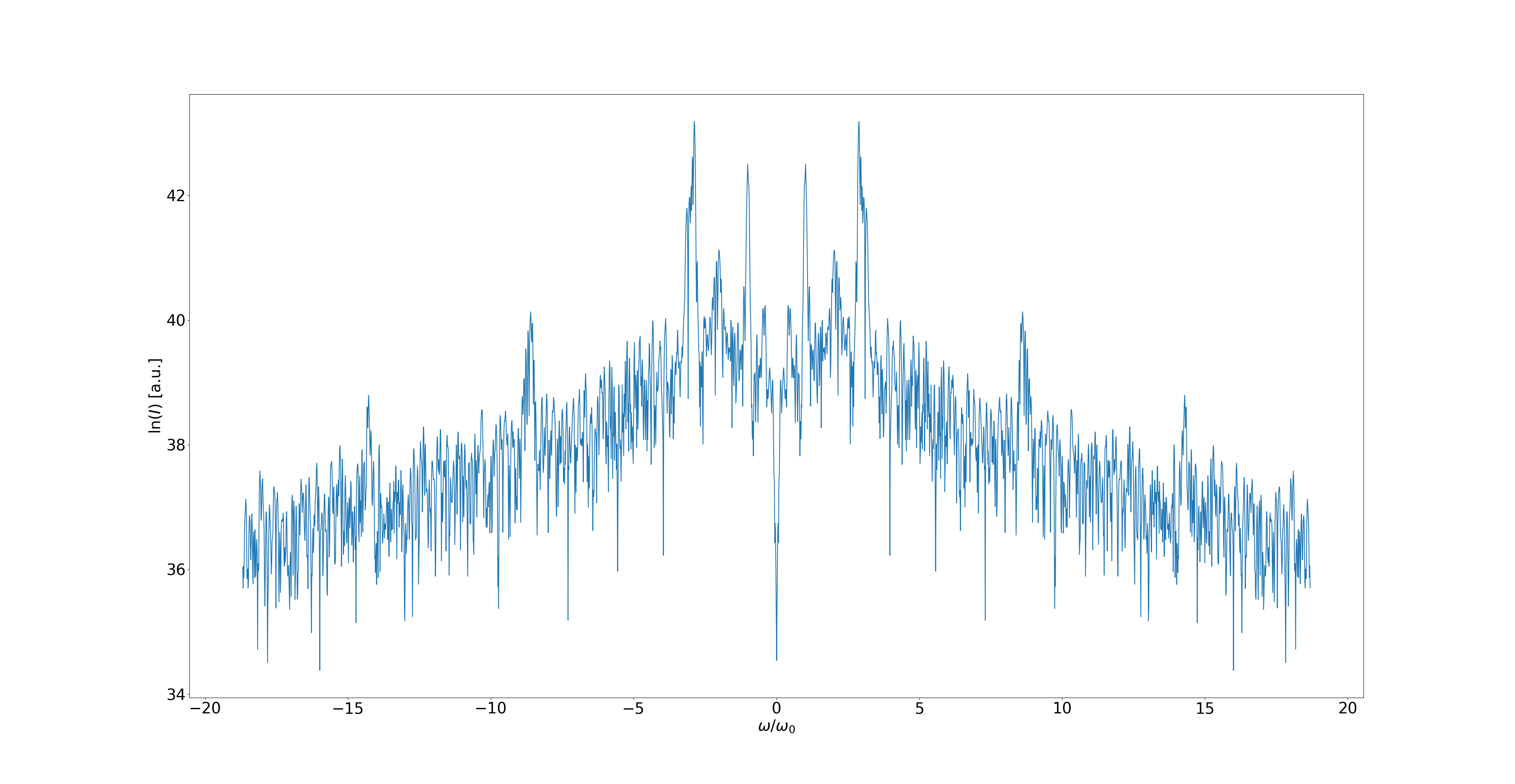}

(e)    \includegraphics[width=0.4\textwidth]{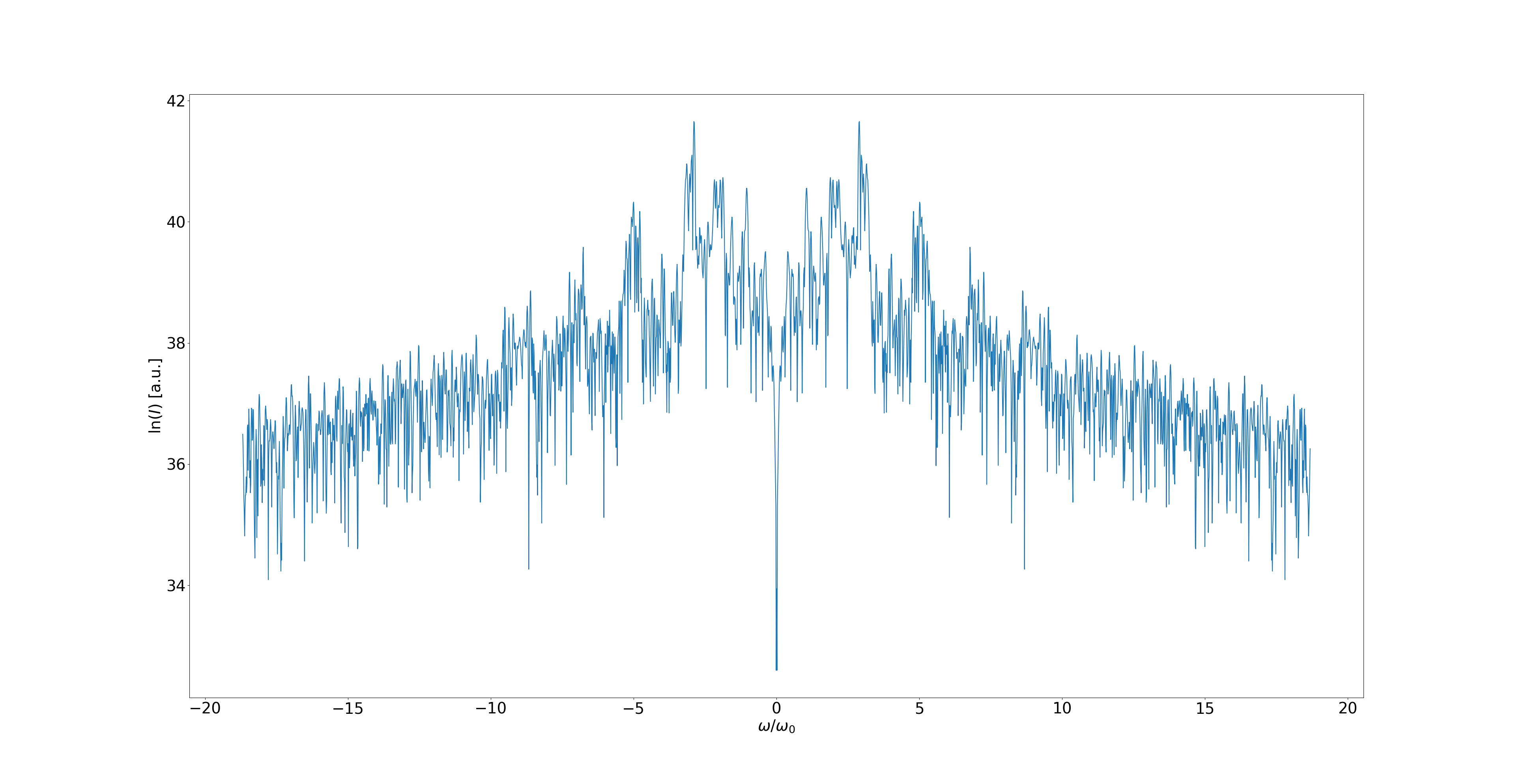}
(f)    \includegraphics[width=0.4\textwidth]{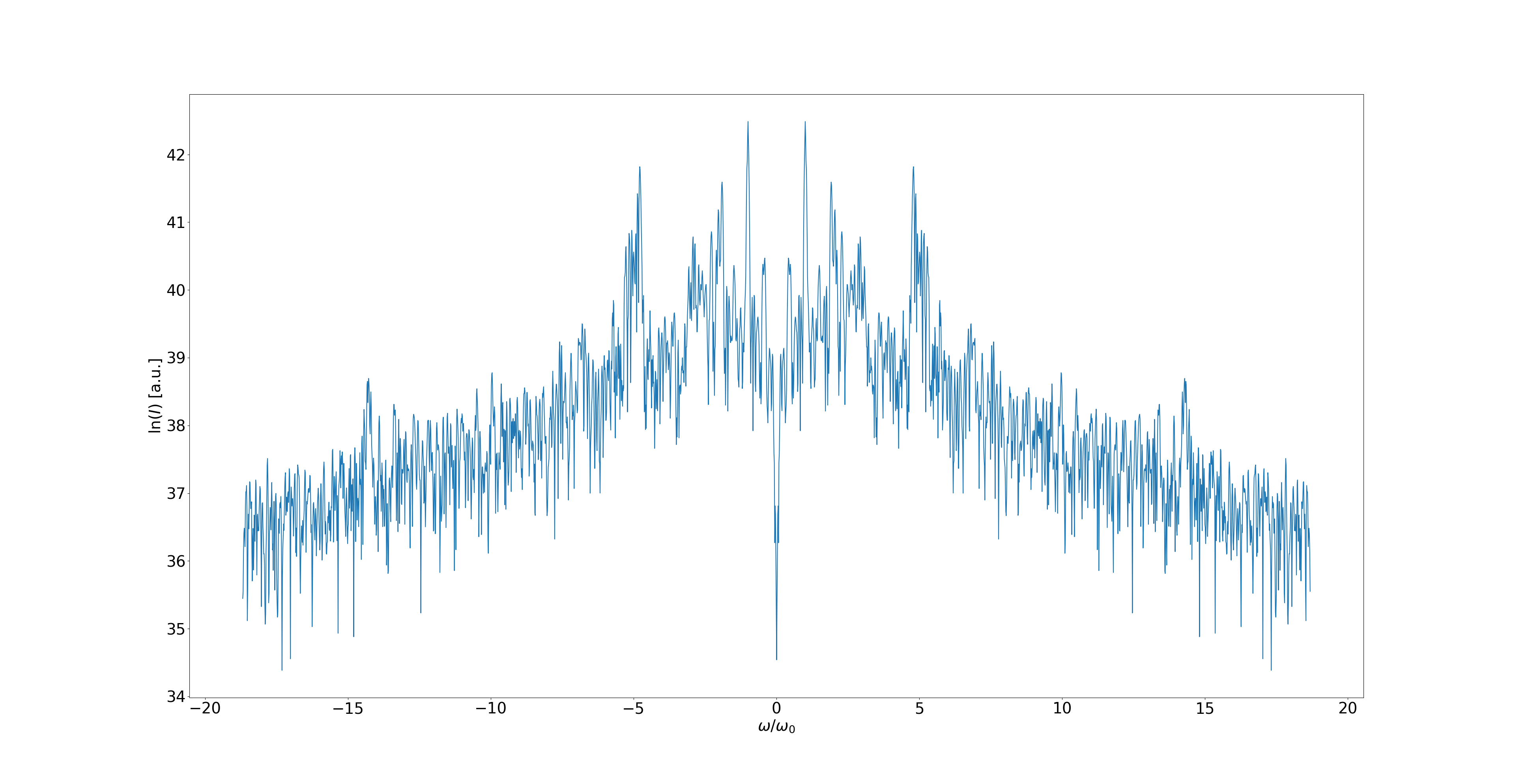}

        \caption{The ``signed frequency'' $\omega/\sigma$ spectrum of the s-polarised radiation corresponding to the line-outs from Figure \ref{figure2}. (a,c,e): results for Gaussian driving beams $\exp(-r^{2})$; (b,d,f): results for super-Gaussian beams $\exp(-r^{16})$. Cuts are taken at $ck_\perp/\omega = \tan(15^\circ)/N$ for $N=1$ (a,b),  $N=3$ (c,d) and $N=5$ (e,f). In each case, spectral peaks are found at $\omega/\sigma = (2n+1)N$, $n\in\Z$. The spectra for siuper-Gaussian driving beams are much clearer than those for Gaussian beams, as discussed in the text. Each image displays a frequency comb with a different line separation, while all images are taken from the results of the same simulation. This proves that a multitude of frequency combs can be obtained from the same laser-target configuration.}
        \label{figure3}
\end{figure*}

The s-polarised harmonics, as given by Eq. (\ref{eq:progression}), all have frequencies and transverse wave numbers that are odd multiples of those of the pump beams. These harmonics do not require the presence of any DC mode to be generated \cite{lichters,trines24}, only the presence of a cubic nonlinear term in the wave equation. This sequence of harmonics can also be found in experiments on HHG in noble gases \cite{hickstein, rego22}, unlike the p-polarised harmonics below, which are more particular to our configuration of two laser pulses hitting a solid surface. The results for this case are shown in Figures \ref{figure1}-\ref{figure3} and summarised below.

\begin{enumerate}
\item The positioning of the peaks in the $(\omega/\sigma, k_\perp/\sigma)$ spectrum is as predicted by Eq. (\ref{eq:progression}).
\item Using a Gaussian pulse led to spectral peaks with a poor shape. This is in line with earlier findings \cite{wikmark,quintard}. For a Gaussian envelope, the radius of the intensity contours shrinks rapidly with increasing intensity, leading to increasing divergence for higher harmonic frequencies.
\item Using a super-Gaussian pulse improved the shape of the spectral peaks significantly. This is in line with findings by Rego \emph{et al.} \cite{rego22}, who used Laguerre-Gaussian pulses with annular envelopes.  For a pulse with either a super-Gaussian or an annular envelope, the radius of the intensity contours does not vary significantly with increasing intensity. The divergence is then proportional to $\lambda_q/R = \lambda_0/(qR)$, decreasing for increasing harmonic order. Thus, the superior divergence properties reported both here and by Rego \emph{et al.} \cite{rego22} are dictated by the envelope of the laser beams employed.
\item The spread in $k_\parallel$ follows from the short duration of the driving pulses and can be reduced by using longer pulses.
\item Wa have taken cross cuts of the two-dimensional harmonic spectrum at $ck_\perp/\omega = \tan(15^\circ)/N$ for $N=1, 3, 5$. These cuts yield harmonic frequency combs with peaks at $(2n+1)N\omega_0$, available from \emph{the same simulation}, demonstrating how versatile our approach is.
\item Peaks ``between'' the laser beams are much higher than those ``outside'' the laser beams. See Ellis \emph{et al.} \cite{ellis} for an explanation: ``sum'' versus ``difference'' harmonic generation, where the ``difference'' process has a higher and lower amplitude than the ``sum'' process.
\end{enumerate}

\subsection{P-polarised harmonics}

\begin{figure*}[ht]
\centering
\includegraphics[width=\textwidth]{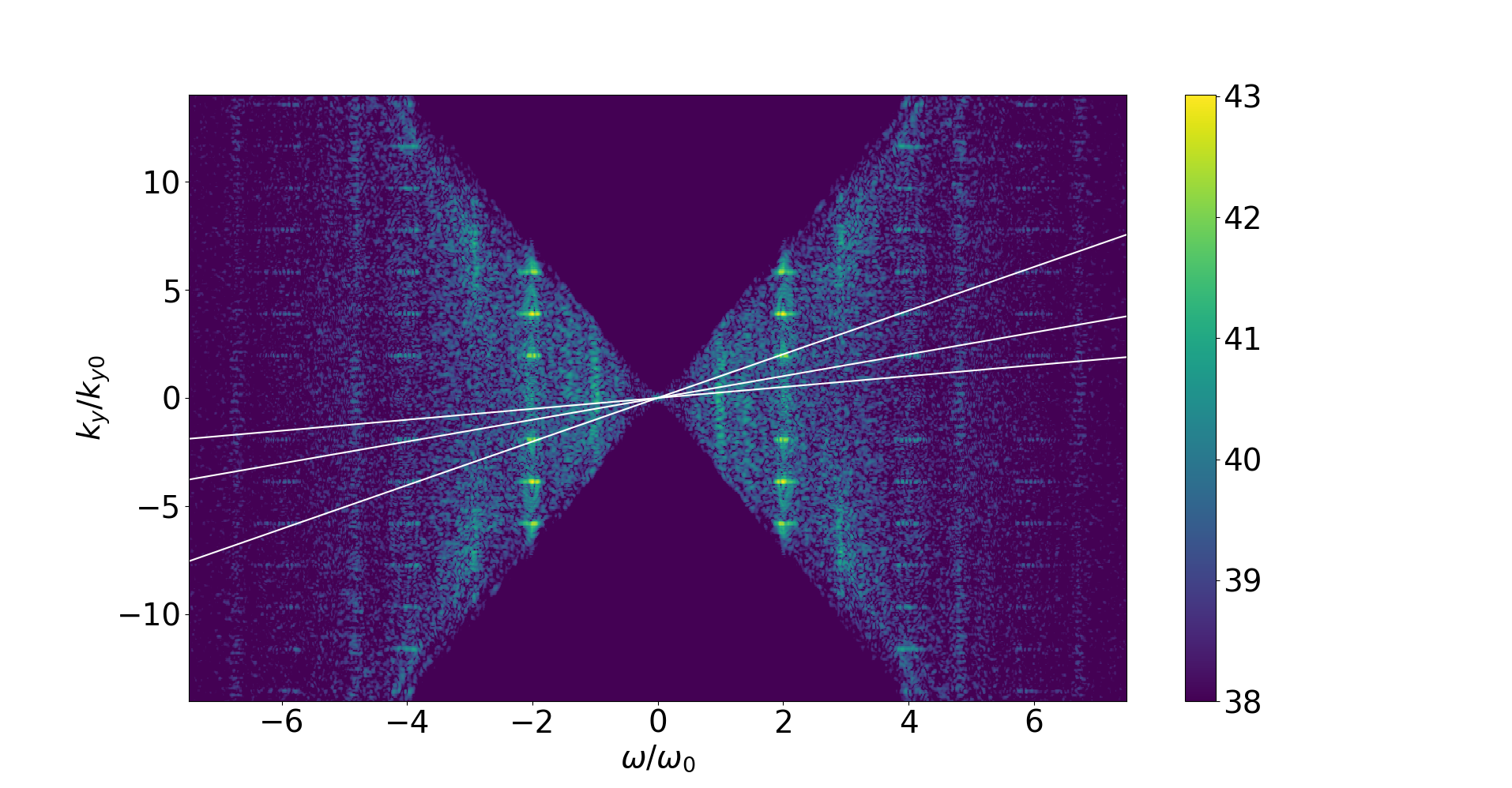}
\caption{The $(\omega/\sigma, k_\perp/\sigma)$ spectrum of the reflected p-polarised harmonic radiation from the super-Gaussian simulation. The positions of the peaks are as predicted by Eq. (\ref{eq:p-progression}). Line-outs are taken at $ck_\perp/\omega = \tan(15^\circ)/N$ for $N=1, 2, 3$, as depicted in the plot (smaller slope means higher $N$). The p-polarised spectrum in this figure is taken from the same simulation as the s-polarised spectrum of Figure \ref{figure3}, once again emphasising that a multitude of frequency combs can be obtained from the same laser-target configuration.}
\label{figure4}
\end{figure*}

\begin{figure*}[ht]
\centering

(a)\includegraphics[width=0.4\textwidth]{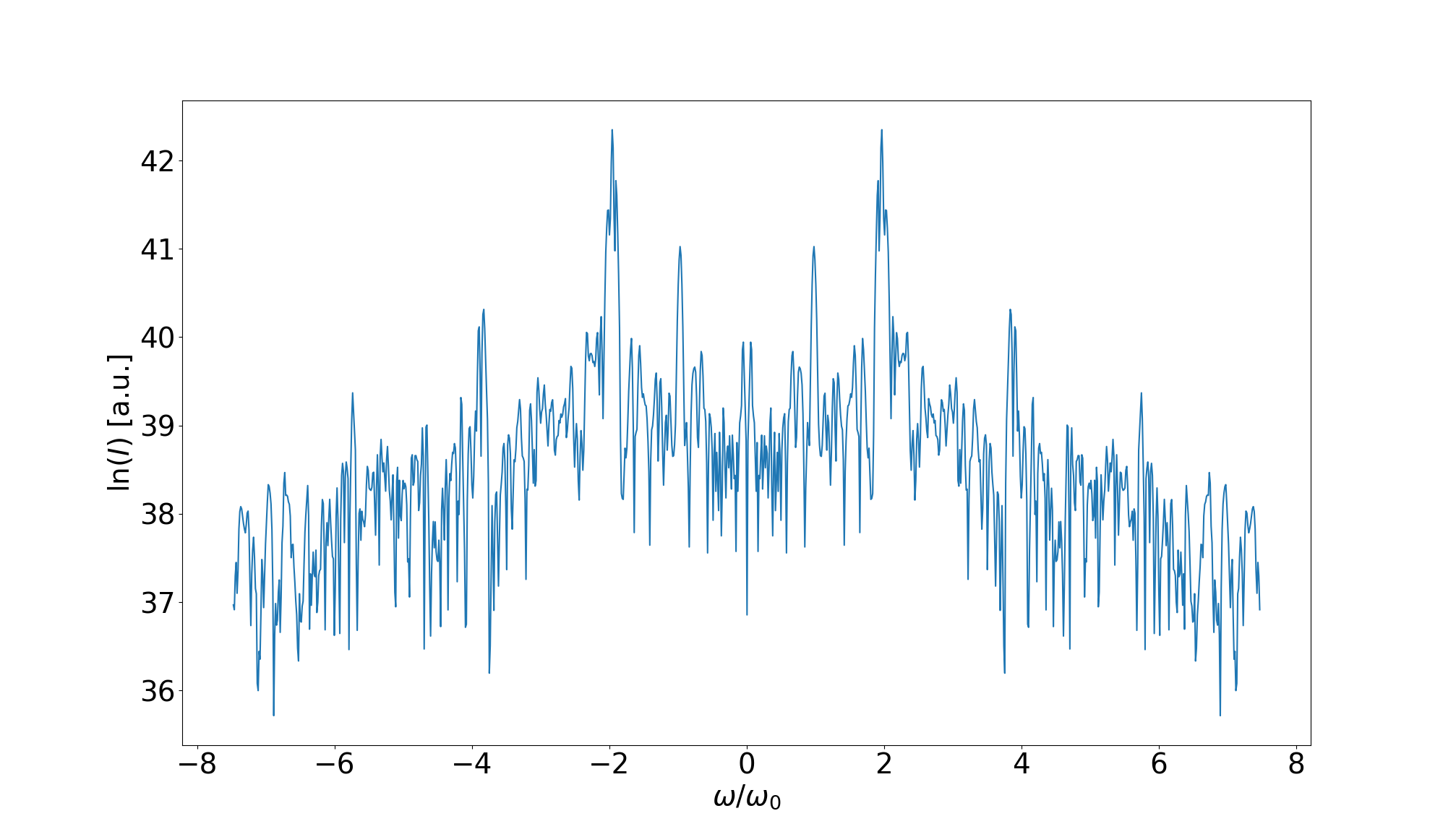}
(b)\includegraphics[width=0.4\textwidth]{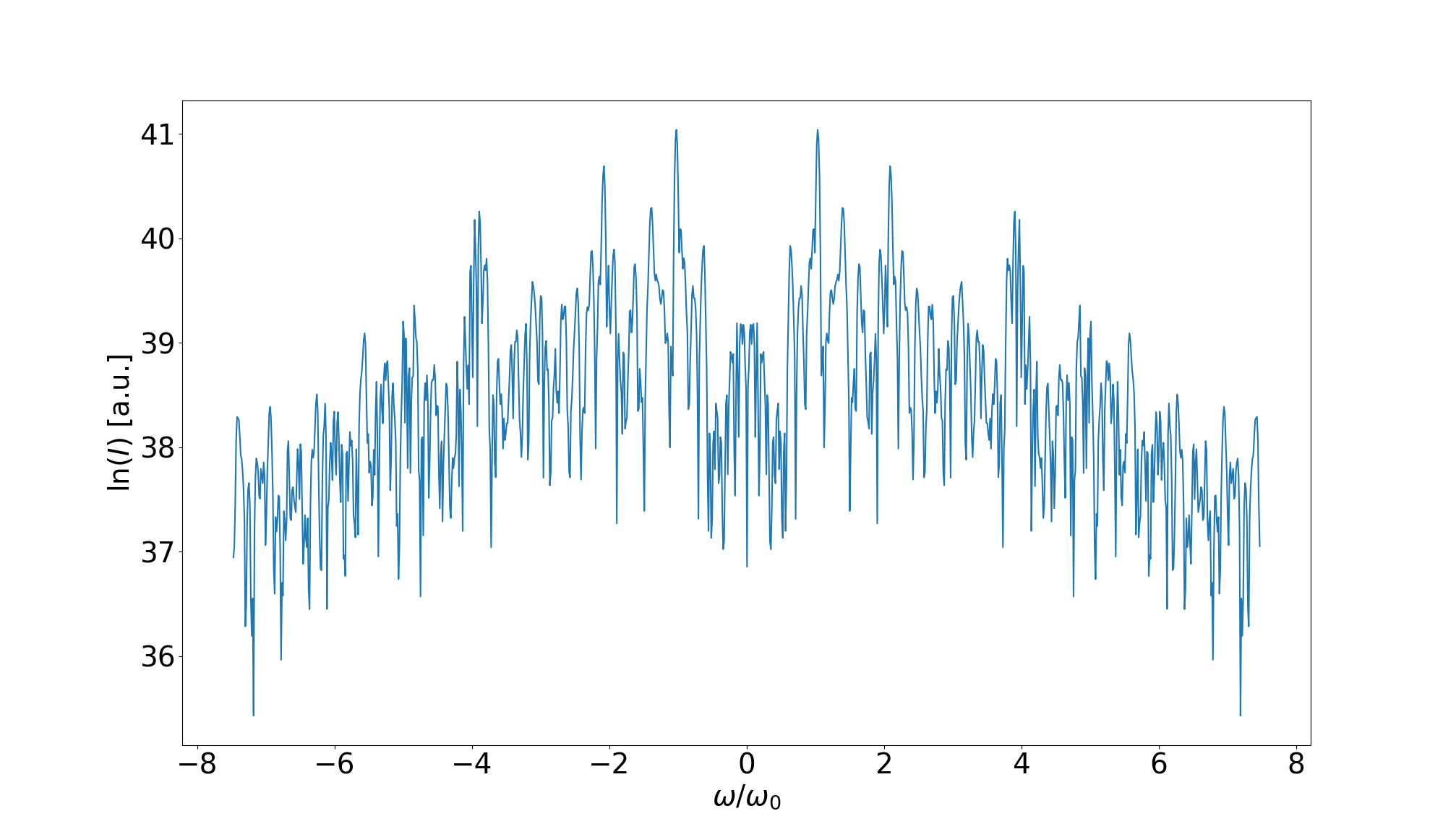}
(c)\includegraphics[width=0.4\textwidth]{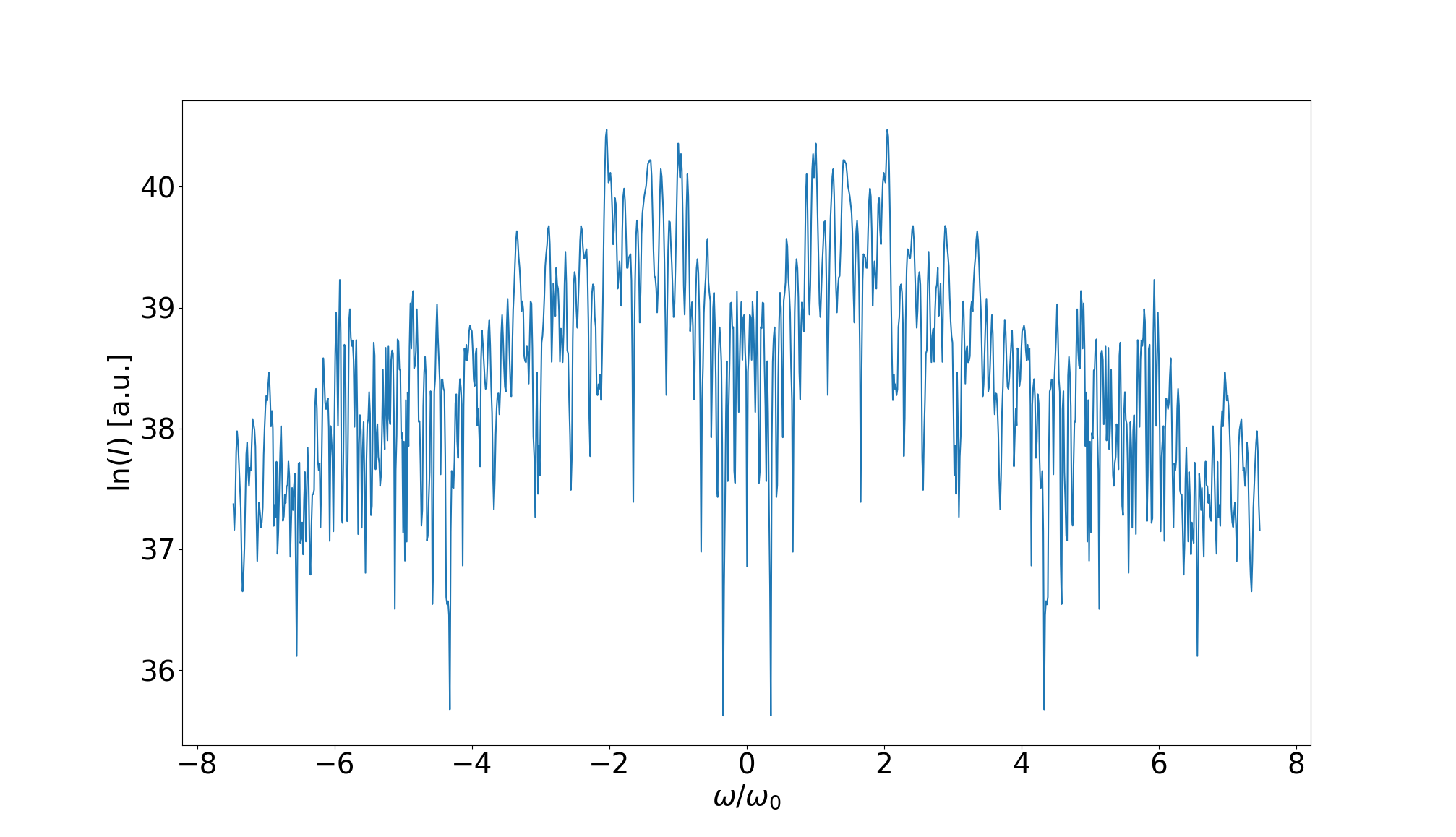}

\caption{Results for the p-polarised light from the super-Gaussian simulation. Line-outs are taken at $ck_\perp/\omega = \tan(15^\circ)/N$ for $N=1, 2, 3$. The ``signed frequency'' $\omega/\sigma$ spectrum of the p-polarised radiation corresponding to the line-outs taken from Figure \ref{figure4} for (a) $N=1$, (b) $N=2$ and (c) $N=3$. In each case, we see that the lowest spectral peaks are at $\omega/(\sigma \omega_0) = \pm N$. While the resolution of our simulation is not sufficient to see a multitude of spectral peaks in the line-outs, the spectrum in Figure \ref{figure4} proves that the underlying principle is sound.}
\label{figure5}
\end{figure*}

From the work by Lichters \emph{et al.} \cite{lichters}, it follows that an s-polarised pump beam, obliquely incident onto a flat surface, will generate even harmonics with p-polarisation. This can be explained as follows. Let $\vec{e}_s$ and $\vec{e}_p$ denote the unit vectors for s- and p-polarisation, with $\vec{e}_s \cdot \vec{e}_p = 0$. Then the pump vector potential is $\vec{A}_0 \propto \vec{e}_s$, while our specific laser-target configuration results in a DC mode $\vec{A}_\mathrm{DC} \propto (m_e c /e) \tan(15^\circ) \vec{e}_p$, and a nonlinear term $\propto A_0^2 \vec{A}_\mathrm{DC}$ in the wave equation, emitting even harmonics with p-polarisation. As discussed above, the harmonic spectrum is given by $(\omega/\sigma, k_\perp/\sigma)_{mn} = (2m,2n)$ for $m,n \in \Z$, and for a viewing angle $\alpha_v$ given by $\tan\alpha_v = k_{\perp 0}/(Nk_{\parallel 0}) = tan(15^\circ)/N$, with $N$ any integer, we preferentially select modes $q\omega_0$ with $q = 2m = (2n)N$, $n\in\N$. Simulation results for the p-polarised light are displayed in Figures \ref{figure4} (full $(\omega/\sigma, k_\perp/\sigma)$ spectrum) and \ref{figure5} (line-outs at $ck_\perp/\omega = \tan(15^\circ)/N$ for $N=1, 2, 3$). We note that these results were obtained from the simulation with the super-Gaussian driving beams discussed above (no separate simulation needed).

\section{Extensions}

Many extensions to our scheme for generating frequency combs are possible. In particular:
\begin{enumerate}
\item Perform HHG in a nonlinear (gaseous) medium, similar to Hickstein \emph{et al.} \cite{hickstein}, rather than reflecting off a solid target. The principle should be the same. Since the light will be crossing the medium rather than scattering back towards the pump lasers, this may be easier to implement in experiments.
\item Use a tilted plane target, so the angles of incidence of the laser beams become unequal. This shifts the grid in the $k_\perp/\sigma$ direction, so it becomes asymmetric with respect to the origin.
\item Use beams with different frequencies. This causes the ``minimal cell'' in the grid to be a parallelogram instead of a rectangle. While non-collinear HHG with $\omega$-$2\omega$ beams has been studied in the past, see e.g. Refs. \cite{bertrand, heyl14, gariepy, zli}, this has not been employed to obtain direction-dependent frequency combs.
\item Use the second transverse dimension: use 4 beams with the crossing angles in the $x$ and $y$ directions slightly different. We diagnose the harmonic radiation via a pinhole rather than a slit. This allows for extra variation in the composition of the frequency comb.
\item Use beams that have already gone through a ``necklace''  HHG process to give them a sparse spectrum, before they cross for a second HHG process.
\item When the harmonic progression has more than 2 dimensions, e.g. 3-D, then one can either choose a 2-D subset on a plane or a 1-D subset on a line. This offers an extra degree of freedom in selecting frequency combs and thus deserves further exploration in a separate publication.
\end{enumerate}

\section{Conclusions}

We have studied HHG in the reflection of two identical high-power laser beams off a plane solid target, where the laser beams have small, opposite angles of incidence with respect to the target normal. The reflected harmonic light does not just contain many frequencies, but also many values for the perpendicular wave number $k_\perp$. For a given oblique observation angle $\alpha$, only those harmonic frequencies that have a component for which $\tan\alpha = k_\perp/k_\parallel$ will be seen. By carefully selecting harmonic light in only a specific direction, a harmonic frequency comb is found where the spacing of the comb depends on the chosen direction. Multiple frequency combs can be obtained from the same simulation, and may thus also be obtained from the same experiment. Also, the s- and p-polarised components of the harmonic light provide different harmonic progressions and thus separate sets of frequency combs, with either odd or even multiples of the pump frequency $\omega_0$. This versatile approach admits multiple extensions, several of which are listed in this paper.

From the results in this work and our previous work \cite{trines24}, we conclude that every 2-D harmonic progression has the potential of a frequency comb. It is mainly a matter of finding the correct viewing angle.

\section*{Acknowledgements}

One of the authors (RT) would like to thank Eva Los and Chris Murphy for useful discussions. Simulations were performed on the Scarf computer cluster at the STFC Rutherford Appleton Laboratory.


\begin{thebibliography}{99}
\bibitem{ferray} M Ferray \emph{et al.}, J. Phys. B: At. Mol. Opt. Phys. {\bf 21}, L31 (1988).
\bibitem{lichters} R. Lichters \emph{et al.}, Phys. Plasmas {\bf 3}, 3425 (1996).
\bibitem{popmin12} T. Popmintchev \emph{et al.}, Science {\bf 336}, 1287 (2012).
\bibitem{popmin15} T. Popmintchev \emph{et al.}, Science {\bf 350}, 1225 (2015).
\bibitem{alon98} O. E. Alon, V. Averbukh and N. Moiseyev, Phys. Rev. Lett. {\bf 80}, 3743 (1998).
\bibitem{bayku21} D. Baykusheva \emph{et al.}, Phys. Rev. A {\bf 103}, 023101 (2021).
\bibitem{rego22} L. Rego \emph{et al.}, Science Advances {\bf 8}, eabj7380 (2022).
\bibitem{trines24} R. Trines \emph{et al.}, Nature Communications {\bf 15}, 6878 (2024).
\bibitem{wikmark} H. Wikmark \emph{et al.}, Proc. Natl. Acad. Sci. U.S.A. {\bf 116}, 4779 (2019).
\bibitem{quintard} L. Quintard, \emph{et al.}, Sci. Adv. {\bf 5}, eaau7175 (2019).

\bibitem{hickstein} D. Hickstein \emph{et al.}, Nature Photonics {\bf 9}, 743 (2015).
\bibitem{fomichev} S. V. Fomichev \emph{et al.}, Laser Physics {\bf 12}, 383 (2002).
\bibitem{bertrand} J. B. Bertrand \emph{et al.}, Phys. Rev. Lett. {\bf 106}, 023001 (2011).
\bibitem{heyl14} C. M. Heyl \emph{et al.}, Phys. Rev. Lett. {\bf 112}, 143902 (2014).
\bibitem{gariepy} G. Gariepy \emph{et al.}, Phys. Rev. Lett. {\bf 113}, 153901 (2014).
\bibitem{zli} Zhengyan Li, \emph{et al.}, Phys. Rev. Lett. {\bf 118}, 033905 (2017).
\bibitem{negro} M. Negro \emph{et al.}, Optics Express {\bf 22}, 29778 (2014).
\bibitem{ellis} J. L. Ellis \emph{et al.}, Optics Express {\bf 25}, 10126 (2017). 

\bibitem{epoch1} T. D. Arber \emph{et al.}, Plasma Phys. Control. Fusion {\bf 57}, 113001 (2015).
\bibitem{epoch2}  The Epoch code is available at \url{https://epochpic.github.io/}

\end{thebibliography}
\end{document}